\documentclass[12pt]{iopart}

\usepackage{iopams,graphicx,xcolor}
\usepackage{enumerate}
\usepackage[breaklinks=true,colorlinks=true,linkcolor=blue,urlcolor=blue,citecolor=blue]{hyperref}

\begin{document}
%Van color is magenta
\def\be{\begin{equation}}
\def\ee{\end{equation}}
\def\bea{\begin{eqnarray}}
\def\eea{\end{eqnarray}}

\title{Experimental proof of Quantum Zeno-assisted Noise Sensing}

\author{Hoang-Van Do$^{1,2}$, Cosimo Lovecchio$^{1,3}$, Ivana Mastroserio$^{1,4,5}$, Nicole Fabbri$^{1,4}$, Francesco S. Cataliotti$^{1,2,4}$, Stefano Gherardini$^{1,2,\ast}$, Matthias M. M\"{u}ller$^{1,6}$, Nicola Dalla Pozza$^{1,2}$, Filippo Caruso$^{1,2}$}

\address{$^1$ LENS and QSTAR, Via N. Carrara 1, I-50019 Sesto Fiorentino, Italy.}
\address{$^2$ Dipartimento di Fisica e Astronomia, Universit\`a degli Studi di Firenze, Sesto Fiorentino, Italy.}
\address{$^3$ Department of Energy, Systems, Territory and Constructions Engineering, University of Pisa, Pisa, Italy.}
\address{$^4$ Istituto Nazionale di Ottica CNR-INO, Firenze, Italy.}
\address{$^5$ Dipartimento di Fisica ``Ettore Pancini'', Universit\`a degli Studi di Napoli Federico II, Napoli, Italy.}
\address{$^6$ Institute of Quantum Control, Peter Gr\"{u}nberg Institut, Forschungszentrum J\"{u}lich, J\"{u}lich, Germany.}
\address{$^\ast$ Corresponding author: gherardini@lens.unifi.it}

\begin{abstract}
In the ideal quantum Zeno effect, repeated quantum projective measurements can freeze the coherent dynamics of a quantum system. However, in the weak quantum Zeno regime, measurement back-actions can allow the sensing of semi-classical field fluctuations. In this regard, we theoretically show how to combine the controlled manipulation of a quantum two-level system, used as a probe, with a sequence of projective measurements to have direct access to the noise correlation function. We experimentally test the effectiveness of the proposed noise sensing method on a properly engineered Bose-Einstein condensate of $^{87}Rb$ atoms realized on an atom chip. We believe that our quantum Zeno-based approach can open a new path towards novel quantum sensing devices.
\end{abstract}

\maketitle

\section{Introduction}

The protection of quantum coherence\,\cite{ZurekRMP2003,StreltsovRMP2017}, or, complementarily, the measurement of environmental effects on a quantum system\,\cite{DegenRMP2017} is of great importance for the development of quantum technologies. In particular, with \emph{quantum sensing} one denotes all techniques that exploit genuine quantum features for the achievement of enhanced measurement performance\,\cite{GiovannettiPRL2006,GiovannettiNATPHOT2011}. Accordingly, the sensing of field fluctuations induced on a quantum system by an external environment, known as \emph{quantum noise sensing}\,\cite{HallPRL2009,ColeNANO2009,Paz-SilvaPRL2014,NorrisPRL2016}, falls within such a framework.

Dynamical decoupling methods have been introduced with the aim of protecting a quantum dynamical evolution from decoherence\,\cite{ViolaPRL1999,KhodjastehPRL2005,UhrigPRL2007,UhrigNJP2008} and to infer specific features of the noise spectrum originating from the environment\,\cite{AlvarezPRL2011,YugePRL2011,BylanderNATPHYS2011,Zwick2016,Paz-SilvaPRA2017,MuellerSciRep2018,PoggialiPRX2018,HernandezPRB2018}. In this regard, quantum-state protection has also been achieved by applying a sequence of quantum projective measurements, often called Zeno measurements, whose action is to freeze the dynamical evolution of the system under observation\,\cite{Misra1977,ItanoPRA1990,KwaitPRL1995,FischerPRL2001}. Moreover, as predicted in\,\cite{PascazioPRL2002,Pascazio2008,SmerziPRL2012,Lerner} and experimentally observed in\,\cite{Schafer2014,Signoles2014}, a quantum system can be also confined in an arbitrary subspace of a larger Hilbert space, according to the so-called quantum Zeno dynamics. However, when the measurement rate is low, repeated observations can enhance the loss of atoms in the measurement subspace and thus allows for the decay of the quantum system. This decrease in survival probability is known as the quantum anti-Zeno effect (QAZE) \cite{Goan,Hanggi,Chaudhry}. The existence of a relationship between dynamical decoupling methods and the application of a sequence of projective measurements for noise spectroscopy is also formally discussed in\,\cite{Sakuldee2019}.

Recently, sequences of Zeno measurements have been studied in the presence of additional stochastic contributions to analyze how the confinement probability of the system to remain in the measured subspace changes with the amount and type of stochasticity\,\cite{GherardiniNJP2016,MuellerAnnalen2017,GherardiniQST2017}. Such a regime is called \emph{weak quantum Zeno} (WQZ) and it has been observed that, although projective measurements could constantly monitor a quantum system, its decay is boosted by the presence of disorder. In general, this process depends both on the spectrum of the noise field and on the coherent control pulses applied to it\,\cite{KofmanNATURE2000,KofmanPRL2001,GordonJPHYSB2007}. Nonetheless, while in the quantum Zeno (QZ) regime the system dynamics is protected against decoherence induced by noise fluctuations, in the WQZ regime, the system may be very sensitive to noisy external fields as e.g.\,magnetic fluctuations\,\cite{MuellerPRA2016,MuellerSciRep2016}.

Here, we propose and demonstrate a novel noise--sensing scheme enabled by the WQZ regime. Specifically, we apply a set of control signals synchronized with projective measurements to enhance the sensitivity of a two--level quantum sensor to a fluctuating resonant field. We validate these theoretical results on a Bose-Einstein condensate (BEC) of Rubidium atoms realized on an atom chip by sensing ad hoc introduced noisy fields with spectra of different nature. For the sake of clarity, note that we do not explicitly consider a physical environment, but we focus on the effects of the external noisy microwave field coupling two hyperfine ground states of the $^{87}$Rb atoms.

The layout of the paper is as follows. In Section\,\ref{section-sensing-scheme}, we introduce our novel noise sensing method, which relies on the combined action of a periodic coherent control field and a sequence of projective measurements into its initial state. We then discuss in Section\,\ref{section-experiment} the experimental procedure that realizes the proposed sensing protocol, by showing its effectiveness in determining the second-order correlation function of the introduced noise fields. Conclusions and outlook follow in Section\,\ref{section-conclusions}.

\section{Sensing scheme in the WQZ regime}
\label{section-sensing-scheme}

Under the effect of a time-dependent Hamiltonian $H(t)$ a state evolves from $|\psi(t_0)\rangle$ into $|\psi(t)\rangle=\mathcal{T}\,e^{-i A(t,t_0)}|\psi(t_0)\rangle$, with $A(t,t_0)=1/\hbar \int_{t_0}^t H(t')dt'$ and $\mathcal{T}$ being the time-ordering operator. If we repeatedly project the system into a given state we can effectively freeze its dynamics in the QZ regime. The latter is obtained when the time interval $\tau$ between consecutive projections is small compared to $\tau_Z=\hbar/\sqrt{{\rm Var}(H)}$, where ${\rm Var}(H)$ is the variance of the Hamiltonian $H(t)$ with respect to the initial state $|\psi(t_0)\rangle$ (see Ref.\,\cite{Pascazio2008}). If, on the contrary, $\tau$ approaches $\tau_Z$, i.e., $\tau=\mathcal{O}(\tau_Z)$, the system is in the WQZ regime\,\cite{MuellerAnnalen2017}, where the survival probability $P$, namely the probability of finding the system in the initial state, decreases quadratically with the expectation value of $A(t+\tau,t)$, always computed with respect to the initial state. When the QZ requirement on the time interval between subsequent measurements is relaxed (i.e., in the WQZ regime), one can then find a condition where the survival probability $P$ is maximally sensitive to an external field. Intuitively, if the projective measurements are too close to each other (pure QZ), the effect of any signal to be measured is inhibited ($P \approx 1$). On the other hand, if projective measurements are too sparse, $P$ decays to zero no matter the presence of an external field.

In the WQZ regime, we can hence devise a method to extract information on a faint noise signal by amplifying its effect with the use of a suitable control field. Let us better illustrate this by introducing the model adopted in our experiment. We consider a quantum system with only two-levels $|0\rangle$ and $|1\rangle$, eigenstates of the computational basis, and $|\psi(0)\rangle=|0\rangle$ as the initial state. We then drive the system with the Hamiltonian
\begin{equation}
H(t)=\frac{\hbar}{2} \left[\Omega_{\rm c}(t) + \Omega_{\rm n}(t)\right]\sigma_x\,,
\end{equation}
where $\Omega_{\rm c}(t)$ is the control field, $\sigma_x$ a Pauli matrix and $\Omega_{\rm n}(t)$ is the unknown stochastic field that we want to characterize. Note that with this definition the Hamiltonian at different time instants commutes. This holds whenever the control and noise fields act on the same operator or, practically, if the dominant component of the noise and the control field are aligned and the other components of the noise are negligible. If we consider additional dephasing, i.e., a contribution acting on $\sigma_z$, it can be shown that this gives only a fourth order contribution to the survival probability and can thus be neglected~\cite{Mueller2019}. While the system evolves, we apply a sequence of $N$ measurements separated by a time interval of $\tau$. By defining
\begin{equation}
   \alpha_j \equiv \int_{{(j-1)\tau}}^{j\tau}\left[\Omega_{\rm c}(t)+\Omega_{\rm n}(t)\right]dt,
\end{equation}
the probability $P$ of finding the system in the initial state after $N$ measurements takes the form:
\begin{equation}\label{eq:eq_3}
    P=\prod_{j=1}^{N}|\langle 0|e^{-(i/\hbar)\alpha_j\sigma_x}|0\rangle|^2=\prod_{j=1}^{N}\cos^2\alpha_{j}.
\end{equation}
Eq.\,(\ref{eq:eq_3}) describes the relative decay of the survival probability $P$, which is given by the product of the survival probabilities in each time interval between two measurements. Then, following Ref.\,\cite{MuellerAnnalen2017}, we can use the second order approximation $\ln\left(\cos^2\alpha_j\right) \approx -\alpha_{j}^{2}$, that is ensured by the condition $\alpha_j^4\ll1$ and defines the WQZ regime. As a result, $P$ can be factorized into three contributions, i.e.,
\begin{equation}
P = \exp\left(-\sum_{j=1}^{N}\alpha_{j}^{2}\right) = P_{\rm c}P_{\rm n}P_{\rm cn},
\end{equation}
where
\begin{eqnarray}
P_{\rm c} & \equiv & \exp\left[-\sum_{j=1}^{N}\left( \int_{{(j-1)\tau}}^{j\tau}\Omega_{\rm c}(t)dt\right)^{2}\right]\label{pc}, \\
P_{\rm n} & \equiv & \exp\left[-\sum_{j=1}^{N}\left( \int_{{(j-1)\tau}}^{j\tau}\Omega_{\rm n}(t)dt\right)^{2}\right] \label{pn},\\
P_{\rm cn} & \equiv & \exp\left[-2
\sum_{j=1}^{N}
\left( \int_{{(j-1)\tau}}^{j\tau}\Omega_{\rm c}(t)dt \right)\left( \int_{{(j-1)\tau}}^{j\tau}\Omega_{\rm n}(t')dt' \right)\right]\label{pcn}.
\end{eqnarray}
$P_{\rm c}$ depends only on the control pulse and thus it can be directly computed, while $P_{\rm n}$ just on the noise field. Under the hypothesis of \emph{weak} noise, i.e., $\sum_{j=1}^{N}\left( \int_{{(j-1)\tau}}^{j\tau}\Omega_{\rm n}(t)dt\right)^{2} \ll 1$, $\ln(P_{\rm n})$ is a second-order term in $\Omega_{n}$ that can be neglected with the result that $P_n \approx 1$, namely $P \approx P_{\rm c}P_{\rm cn}$. Finally, $P_{\rm cn}$ is a cross-term of noise and control containing all the interesting information on the spectral properties of $\Omega_{\rm n}(t)$. In other words, a measurement of the survival probability $P$ gives direct access to $P_{\rm cn} \approx P/P_{\rm c}$. Note that, in the absence of the control field, the validity of the weak noise condition implies that the system is practically operating in the \emph{strong} (stochastic) QZ regime\,\cite{MuellerAnnalen2017} and the survival probability $P$ deviates from the value predicted by the theory of standard (non-stochastic) Zeno physics only by a small amount (due to fluctuations), but is still very close to 1. If we turn on the control field (which is stronger than the noise field), the probe operates in the weak Zeno regime, so that the survival probability slowly decays and the decay can be described by the second-order approximations made above. Also, the slope of the decay reveals information on the interplay of noise and control. To make accessible different portions of the noise in the frequency domain, we resort to the action of a set of properly designed control pulses that filter out these frequency contributions.

Now, let us rewrite equation (\ref{pcn}) as
\begin{equation}
\ln{(P_{\rm cn})}=-2\int_{0}^{N\tau}\widetilde\Omega_{c}(t)\Omega_{n}(t)dt\,,
\label{logPoverPcc}
\end{equation}
where we have defined
\begin{equation}
\widetilde\Omega_{\rm c}(t) \equiv \sum_{j=1}^{N}\left( \int_{{(j-1)\tau}}^{j\tau}\Omega_{\rm c}(t')dt'\right)\mathcal{W}_j(t),
\end{equation}
with the rectangular window function $\mathcal{W}_j(t)$ equal to $1$ for $(j-1)\tau\leq t<j\tau$ and $0$ otherwise. Eq.\,(\ref{logPoverPcc}) naturally suggests the possibility of sampling the noise with a filter function dictated by modulating the control field. Notice that, being $\widetilde\Omega_c(t)$ a piecewise function constant between each couple of measurements, one can naturally choose the periodicity of the control field $\Omega_c(t)$ to be a multiple of $\tau$. Thus, by averaging $\ln^{2}{\left(P_{\rm cn}\right)}$ over different experimental realizations, we can directly obtain the \emph{second-order correlation function} $\chi_{N}^{(2)}$, which quantifies the correlations between noise and control. In particular, it can be proven that
\begin{equation}
\frac{1}{4}\langle\ln^{2}{\left(P_{\rm cn}\right)}\rangle = \int_{0}^{N\tau}\int_{0}^{N\tau}\widetilde\Omega_{\rm c}(t)\widetilde\Omega_{\rm c}(t')\langle\Omega_{\rm n}(t)\Omega_{\rm n}(t')\rangle\,dtdt' \equiv \chi_{N}^{(2)},
\label{chi}
\end{equation}
where we have assumed, without loss of generality, that the average of the noise (a real, stationary stochastic process) over a sufficiently large number of realizations is zero in each time instant $t$. As a matter of fact, a noise with a non-zero average can be mathematically modelled as a part of the control field. From Eq.\,(\ref{chi}) we can have access to $\langle\Omega_{\rm n}(t)\Omega_{\rm n}(t')\rangle$, which is the noise autocorrelation function\,\cite{DegenRMP2017}. In this regard, if we introduce the noise power spectral density $S(\omega)$ as the Fourier transform of the autocorrelation function, we can express $\chi_{N}^{(2)}$ in the frequency domain and thus write
\begin{equation}\label{eq:overlap_spectrum_filter}
\chi_{N}^{(2)} = \int_{0}^\infty S(\omega)F(\omega)d\omega\,,
\end{equation}
where the \emph{filter function} $F(\omega)\equiv\frac{1}{2\pi}\left|\int_0^{N\tau}\widetilde\Omega_{\rm c}(t)e^{-i\omega t}dt\right|^2$ is the absolute square of the normalized control field short--time Fourier transform. Once again, it is worth noting that Eq.\,(\ref{eq:overlap_spectrum_filter}) holds true only if the noise is a stationary process and thus its power spectral density does not change over time. Therefore, by measuring $\chi_{N}^{(2)}$ for different choices of the control pulse $\widetilde{\Omega}_{\rm c}(t)$, one can infer the power spectral density $S(\omega)$ of the noise in different frequency regimes.

To summarize, the above considerations lead us to the following \emph{noise sensing protocol}:

\begin{enumerate}[(I)]
\item Initialize the two-level system in $|0\rangle$
\item Apply the measurement projector $|0\rangle\langle 0|$ $N$ times with a repetition rate $1/\tau$, driving the system with a control field of period $2\tau$
\item Measure the probability $P$ for the system to remain in $|0\rangle$ after $N$ measurements
\item For a set of values $\tau_{\rm min}< \tau< \tau_{\rm max}$\,, steps (I)-(III) are repeated $Q$ times for each chosen value of $\tau$.
\end{enumerate}

Notice that, by assuming that the value of $Q$ is large enough, the average of the noise $P_n$ at the final time instant $N\tau$ of the procedure is practically vanishing. From this assumption, one finds that the average of $\ln(P_{\rm cn})$ over $Q$ is zero ($\langle\ln P_{\rm cn}\rangle=0$), provided that the control and the noise are uncorrelated signals. Thus, being $\langle P\rangle=P_c$, the second-order correlation function of the noise can be written as
\begin{equation}
\chi^{(2)}_N={\rm Var}\left[\ln\left(\frac{P}{\langle P\rangle}\right)\right],
\label{chivar}
\end{equation}
showing that $\chi^{(2)}_N$ is simply related to the variance of the logarithm of the survival probability $P$. In conclusion, the proposed protocol gives access to the noise second-order correlation function in a bandwidth $\left[\frac{1}{2\tau_{\rm max}}\,,\frac{1}{2\tau_{\rm min}}\right]$ with a frequency resolution of $\frac{1}{N\tau}$.

As a final remark, we want to make a brief comment on the relation between our method and pulsed dynamical decoupling spectroscopy. The latter is based on the coherent control of the quantum system used as a probe, and enables the inference of noise field features by applying a sequence of $\pi$ pulses. These $\pi$ pulses, similarly to our Zeno projection measurements, give a basic frequency grid. However, on top of this, we can also modulate the control pulse $\Omega_c(t)$, so that we can tune the protocol to different noise strength. The key idea here is, that more frequent interactions slow down the decay, while the modulation of $\Omega_c(t)$ (or $\widetilde\Omega_c(t)$) determines the frequency of the spectral density that we want to filter. In the dynamical decoupling approach there is no immediate way to separate these two types of manipulations.
However, a fair and in-depth comparison of the two approaches is beyond the scope of this paper and the preferred protocol will probably depend on the noise field and the probe system.

\section{Experiment}
\label{section-experiment}

\begin{figure}[h!]
\begin{center}
\includegraphics[width=0.875\textwidth]{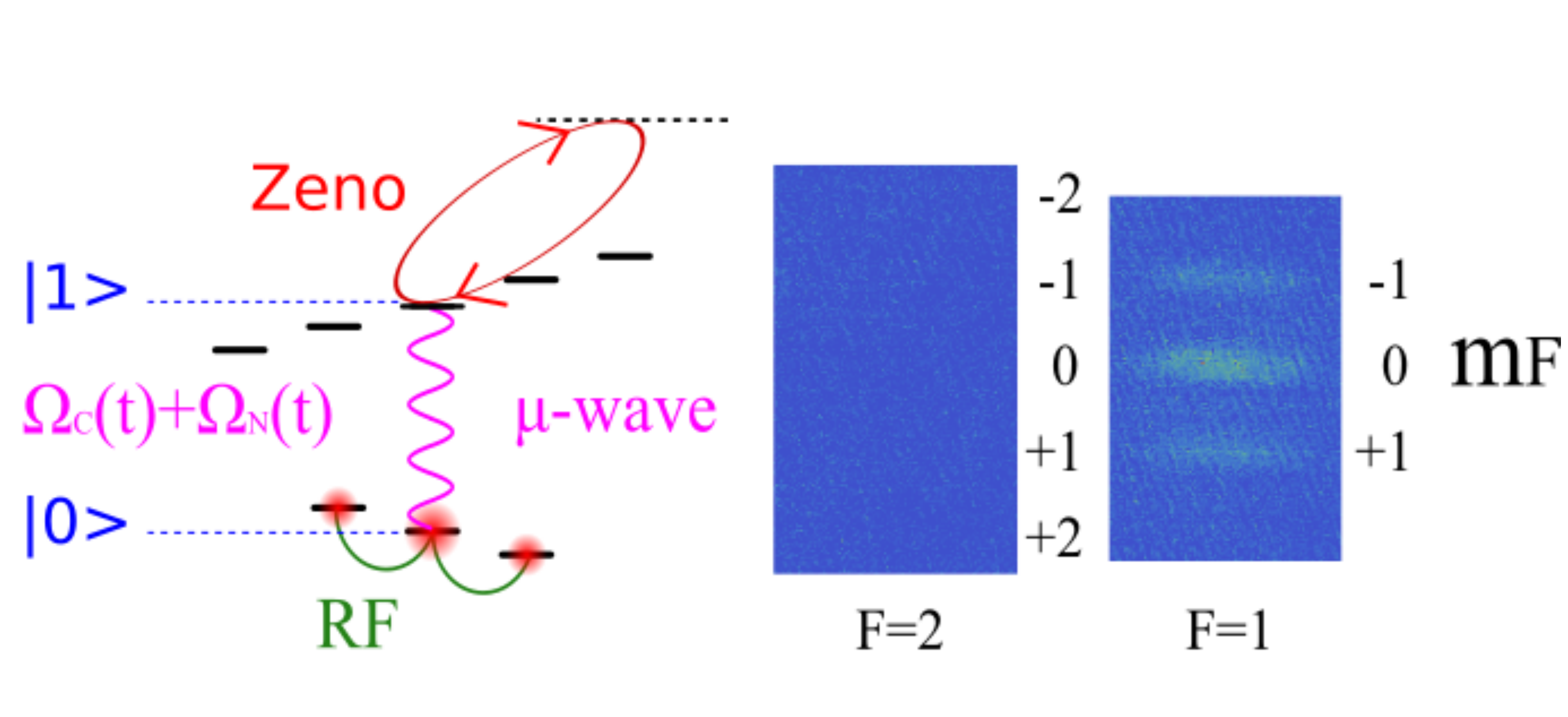}
\caption{Scheme of the level structure of $^{87}$Rb in the presence of a weak homogeneous magnetic field. A quasi resonant microwave field is used to couple different hyperfine levels (pink line), and an optical field (red line) to realize the QZ dynamics by coupling the system to an external excited state. The atoms are initially distributed, by means of a radio frequency field (green line), in the three magnetic sub-levels of $|F=1\rangle$ with half of the population in the $|0\rangle$ state and the other half of the population in states $|F=1,m_F=\pm1\rangle$.}
\label{fig:Fig1}
\end{center}
\end{figure}

We test the proposed sensing protocol on a BEC of $^{87}$Rb atoms, realized on an atom chip. The two--level system is given by the two hyperfine states $|F=1,m_F=0\rangle\equiv|0\rangle$ and $|F=2,m_F=0\rangle\equiv|1\rangle$ of $^{87}$Rb. The conceptual scheme of the experiment is sketched in Fig.\,\ref{fig:Fig1}.

We initially prepare a BEC of typically $\sim 6\times 10^4$ atoms in the $|F=2,m_F=+2\rangle$ magnetic hyperfine sub-level. The trapping magnetic fields and the radio frequency (RF) fields used for evaporating and subsequently manipulating the atoms are produced by conductors integrated into the chip\,\cite{Petrovic}. To suppress interactions on relevant timescales, after reaching quantum degeneracy, the gas is released from the magnetic trap and expanded for a time-of-flight of 0.7~ms, strongly reducing inter-atomic collisions. The magnetic degeneracy of the hyperfine levels is then lifted by applying a homogeneous and constant magnetic field of 6.179~G. Thanks to the opposite sign of the Land\'e factors in the two hyperfine ground levels, this leads to different effective 2-level systems in the $|F=1\rangle\rightarrow|F=2\rangle$ microwave transition. A microwave field with a frequency around 6.834~GHz is used to drive the transition between $|F=2,m_F=0\rangle\equiv|1\rangle$ and $|F=1,m_F=0\rangle\equiv|0\rangle$, while a quasi-resonant RF field is used to couple the sub-levels $m_F=\lbrace+2,+1,0,-1,-2 \rbrace$ of $|F=2\rangle$ and $m_F=\lbrace +1,0,-1 \rbrace $ of $|F=1\rangle$.

A frequency-modulated RF pulse, designed with an optimal control strategy\,\cite{LovecchioPRA2016}, transfers all the atoms from the initial $|F=2,m_F=+2\rangle$ to the $|F=2,m_F=0\rangle$ state. Subsequently a microwave $\pi$ pulse transfers the atoms in $|F=1,m_F=0\rangle$. With this procedure, the system is initialized in the $|0\rangle$ state.

Since the number of atoms in the BEC fluctuates from one realization to another, after the preparation of the atoms in the $|0\rangle$ state, we apply an RF $\pi/2$ pulse placing half of the population in states $|F=1,m_F=\pm1\rangle$. Atoms in these states are not affected during the experiment, so they can be used as a reference for normalization. Note that, since we are working with identical and non-interacting atoms, the relative atomic population of the three sublevels effectively describes the probability for each atom to be found in one of its three sublevels.

During the subsequent evolution, we illuminate the atoms with a light pulse of $\tau_{\rm m} = 0.6\rm\,\mu s$ duration, resonant with the $|F=2\rangle\rightarrow|F=3\rangle$ component of the Rubidium $D2$ line. Note that from the excited state $|F=3\rangle$ atoms will immediately decay outside the condensate and will, therefore, be lost. On condition that the atoms are still in the BEC, this effectively realizes the projective measurement into the ground state, mathematically described by the projector $|0\rangle\!\langle 0|$.

The number of atoms in each of the internal states is measured by applying a Stern-Gerlach method. As a matter of fact, after the noise sensing procedure, we let the atoms fall in the presence of an inhomogeneous magnetic field applied along the quantization axis. This magnetic field spatially separates the different magnetic sublevels $m_F$. At the end, a standard destructive absorption imaging sequence on the $D2$ line is executed\,\cite{Ketterle} -- see also the right side of Fig.\,\ref{fig:Fig1}. Since the hyperfine states are separated by a frequency much larger than the $D2$ linewidth, they can be easily distinguished, by performing two consecutive imaging procedures with light resonant with the two hyperfine components.

All our experimental points are the average of $Q=14$ realizations with error bars coming from the data standard deviation. This number of realizations is chosen as a trade--off between the maximization of statistics and the minimization of experimental fluctuations over a long period.

\subsection*{Protocol implementation and Results}

In our noise sensing protocol, we are supposed to drive the $|0\rangle\rightarrow|1\rangle$ transition. To this aim, we use a resonant microwave field with an amplitude that is modulated by a zero average square-wave with $2\tau$ period, i.e.,
\begin{equation}
\Omega_{\rm c}(t) = \sum_{j=1}^{N}(-1)^j\Omega_0\mathcal{W}_j(t),
\end{equation}
with $\Omega_0=2\pi \times 43.3\,\rm kHz$. Then, in correspondence of each switching of the control Hamiltonian, i.e., at $t=j\tau$, a projective measurement is applied. Being limited by the stability of the light pulse generator, the number of projective measurements is chosen equal to $N=18$. In the experiment also the minimum repetition time $\tau_{\rm min}>2 \times 0.6\rm\,\mu s$ is constrained by the duration $\tau_{\rm m}$ of the light pulse implementing the projector. Moreover, we need to have $N\tau_{\rm max}<100\,\rm\mu s$ because of experimental decoherence.

Following the protocol, the noise field used in the experiment is represented by a resonant microwave with a time-dependent amplitude. As preliminary result needful to properly tune the experimental parameters, we apply a modulation with a single-frequency component i.e., the sinusoidal signal $\Omega_n(t)=\Omega_{\rm n0}\sin\left(\omega_{N} t+\phi\right)$ with random phase $\phi$, $\Omega_{\rm n0}=2\pi \times 12\rm\,kHz$ and $\omega_{N}=2\pi \times 167\rm\,kHz$\footnote{This value, which denotes the single-frequency component of the sinusoidal signal, will be also the central frequency of all our noise power spectral densities. It has been chosen so that the sensitivity of $\mathcal{P}$ with respect to the action of the control is maximized.}.
\begin{figure}[t!]
\begin{center}
\includegraphics[width=0.78\textwidth]{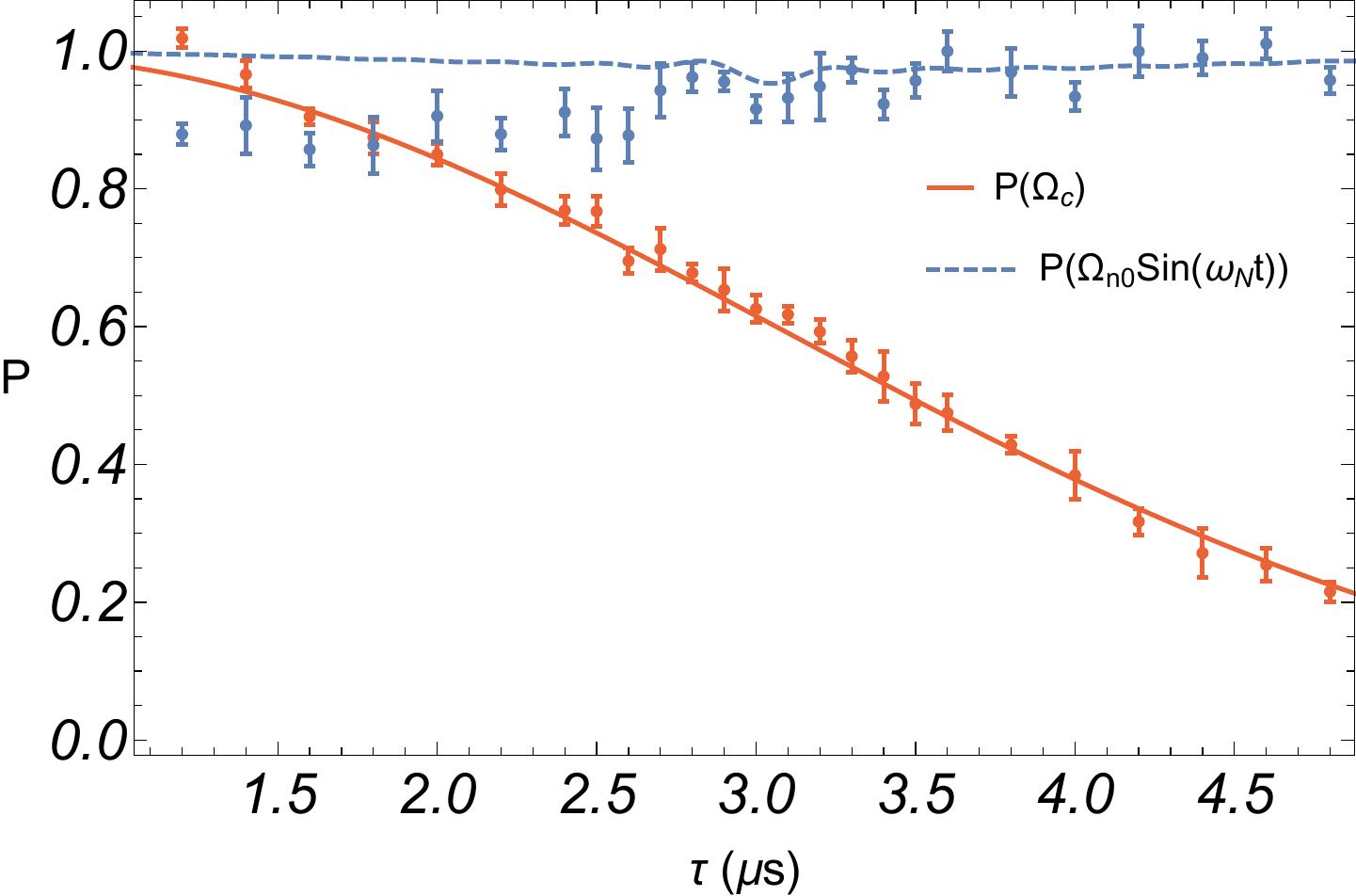}
\caption{Survival probability $\mathcal{P}$ as a function of $\tau$, respectively, recorded in the presence of only the control field $\Omega_c$ (in red) and only the sinusoidal signal $\Omega_n = \Omega_{\rm n0}\sin\left(\omega_{N}t\right)$ (in blue). The dots are the experimental results, while the continuous and dashed lines show the dynamics of the system as predicted by Eqs.\,(\ref{pc}) and (\ref{pn}). Notice that since also the RF pulse used for normalization can produce a rotation slightly drifted away from $\pi/2$ from one realization to another, it is possible to experimentally measure values of the survival probability higher than one. Also, the dip in the blue dashed line appears at $\tau=3\mu s$ as predicted by Eqs.\,(\ref{eq:eq_3}) and (\ref{pn}), that is, when the noise period $2 \tau$ equals the control period $T=\frac{2\pi}{\omega} = 6 \mu s$ and their phases are synchronized.}
\label{fig:fig_Pc}
\end{center}
\end{figure}
However, in Fig.\,\ref{fig:fig_Pc} we first consider the \emph{deterministic} scenario given by imposing $\phi=0$. In this regard, in Fig.\,\ref{fig:fig_Pc} we report in red the experimental survival probability $\mathcal{P}$, i.e., the probability for an atom to remain in the initial state $|0\rangle$, as a function of $\tau$. In this case, $\mathcal{P}=P_c$ and exponentially decreases with the square of $\tau$, in perfect agreement with Eq.\,(\ref{pc}) (continuous red line). Instead, the blue curve in Fig.\,\ref{fig:fig_Pc} is the survival probability $\mathcal{P}$ as a function of $\tau$ with the only presence of the sinusoidal signal $\Omega_n = \Omega_{\rm n0}\sin\left(\omega_{N} t\right)$. In this case, $\mathcal{P}$ is compatible with $1$ within the experimental errors, thus confirming the hypothesis that $P_n \approx 1$, at least in its average value.

\begin{figure}[t!]
\begin{center}
\includegraphics[width=0.8\textwidth]{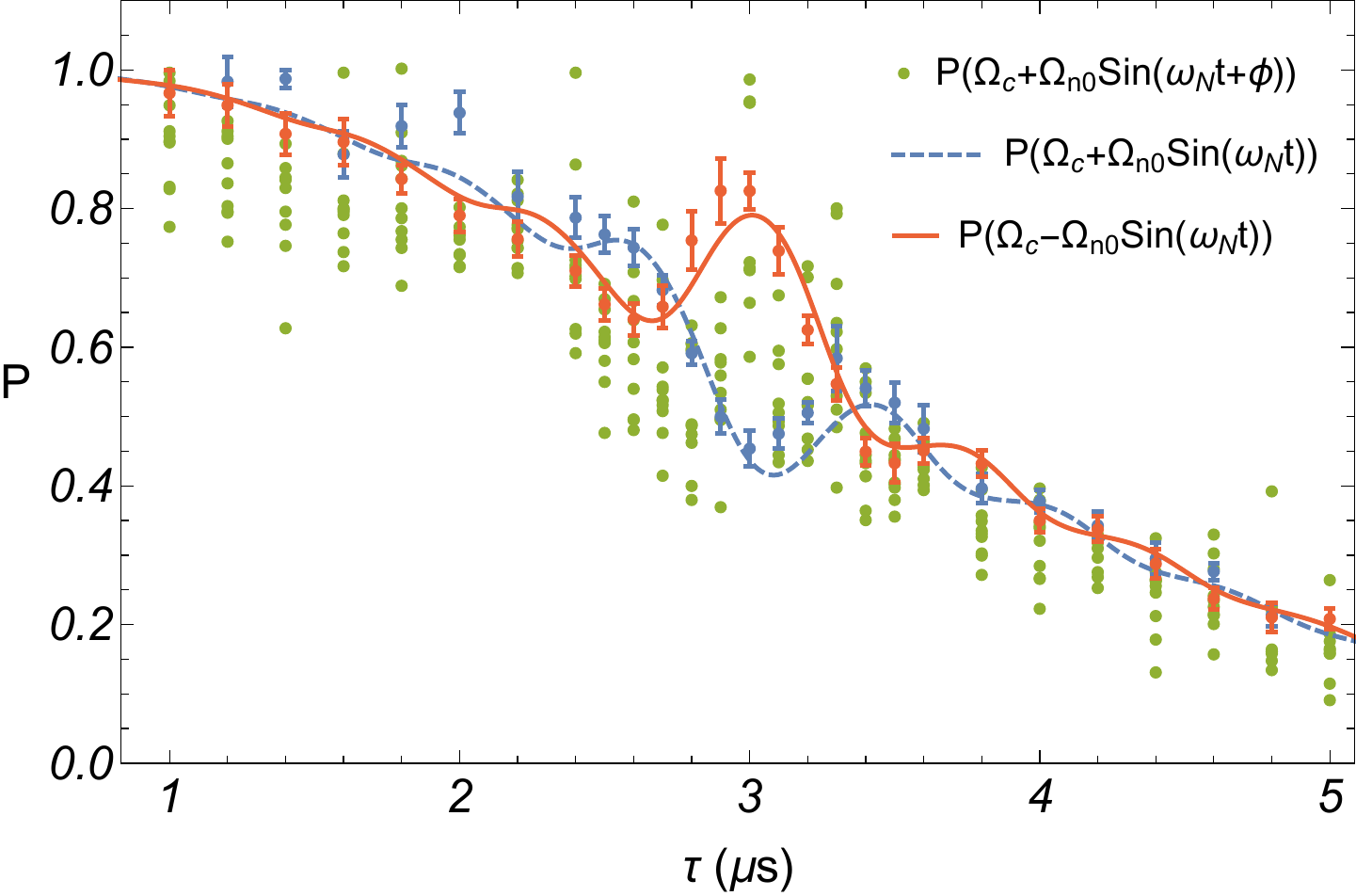}
\caption{Survival probability $\mathcal{P}$ as a function of $\tau$ in the presence of the control field, denoted as $\Omega_c$ in the legend of the figure, and different noise signals $\Omega_n$.  In red: sinusoidal signal without additional phase; in blue: sinusoidal signal with opposite phase. The dots are the experimental results, while the continuous and dashed lines denote the prediction of Eq.\,(\ref{pcn}). Moreover, the green dots show the fluctuations recorded for a sinusoidal noise with a random phase, i.e., a single frequency component noise.}
\label{fig:sin}
\end{center}
\end{figure}

The effect of $\Omega_n$ is amplified by the control field, as shown in Fig.\,\ref{fig:sin}. It exemplifies the working principle of our sensing method. Indeed, by switching the phase $\phi$, the survival probability is either enhanced for $\phi=0$ or decreased for $\phi=\pi$. As a consequence, when $\phi$ is randomly chosen from a uniform probability distribution in the range $[0,2\pi]$, the average of the survival probability coincides with the one due to the control field alone, i.e., $\langle\mathcal{P}\rangle=P_c$. Meanwhile, its variance $\sqrt{{\rm Var}\left(\mathcal{P}\right)}$ is maximized when the repetition rate corresponds to half of the frequency of the noise component, i.e., $1/\tau\sim\omega/\pi$.

\begin{figure}[h!]
\begin{center}
\includegraphics[width=0.8\textwidth]{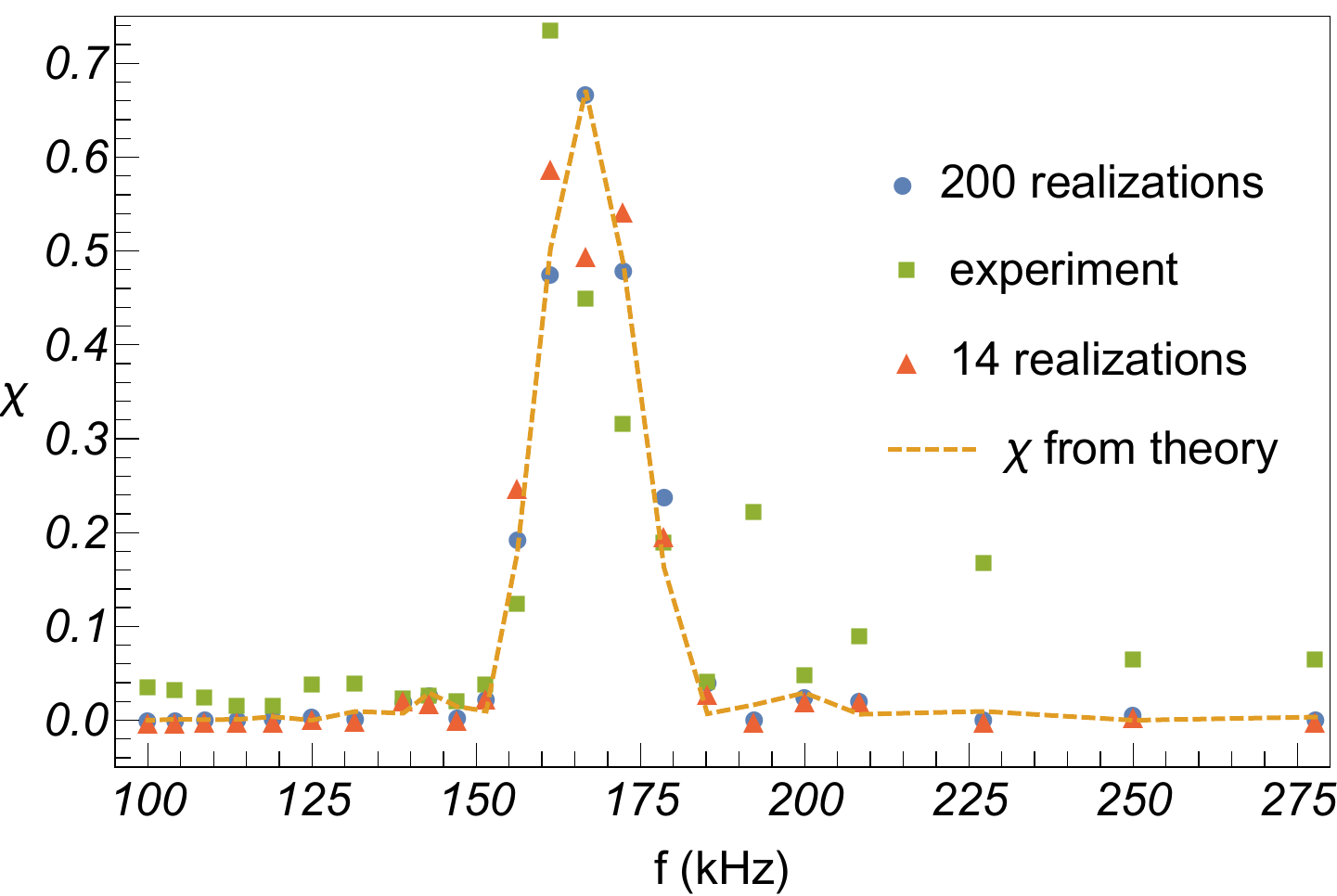}
\caption{Second-order correlation function $\chi^{(2)}_N$ as a function of frequency for a finite single frequency component noise with a random phase. For $Q=14$ the green squares show the experimental results, while the dashed yellow line denotes the theoretical curve obtained by evaluating Eq.\,(\ref{eq:overlap_spectrum_filter}). Simulating the time dynamics, we attain numerical values both for $Q=14$ (red triangles) and $Q=200$ (blue circles).}
\label{fig:chisin}
\end{center}
\end{figure}

In Fig.\,\ref{fig:chisin}, we report the experimental, numerical and theoretical $\chi^{(2)}_N$ for the same number of realizations $Q=14$ of the protocol, with a remarkably good agreement.
By increasing $Q$, the numerical data, obtained by simulating the time dynamics for different realizations of the noise, approach the theoretical curve -- calculated from Eq.\,(\ref{eq:overlap_spectrum_filter}) -- that, in the case of a single noise component for a finite time window, is a sinc$^2$ function. Notice that, since in the experiment we have the possibility of switching off the noise, thus gaining direct access to $P_c$, we use this value instead of the average of $\mathcal{P}$ over 14 realizations.

\begin{figure}[h!]
\begin{center}
\includegraphics[width=0.825\textwidth]{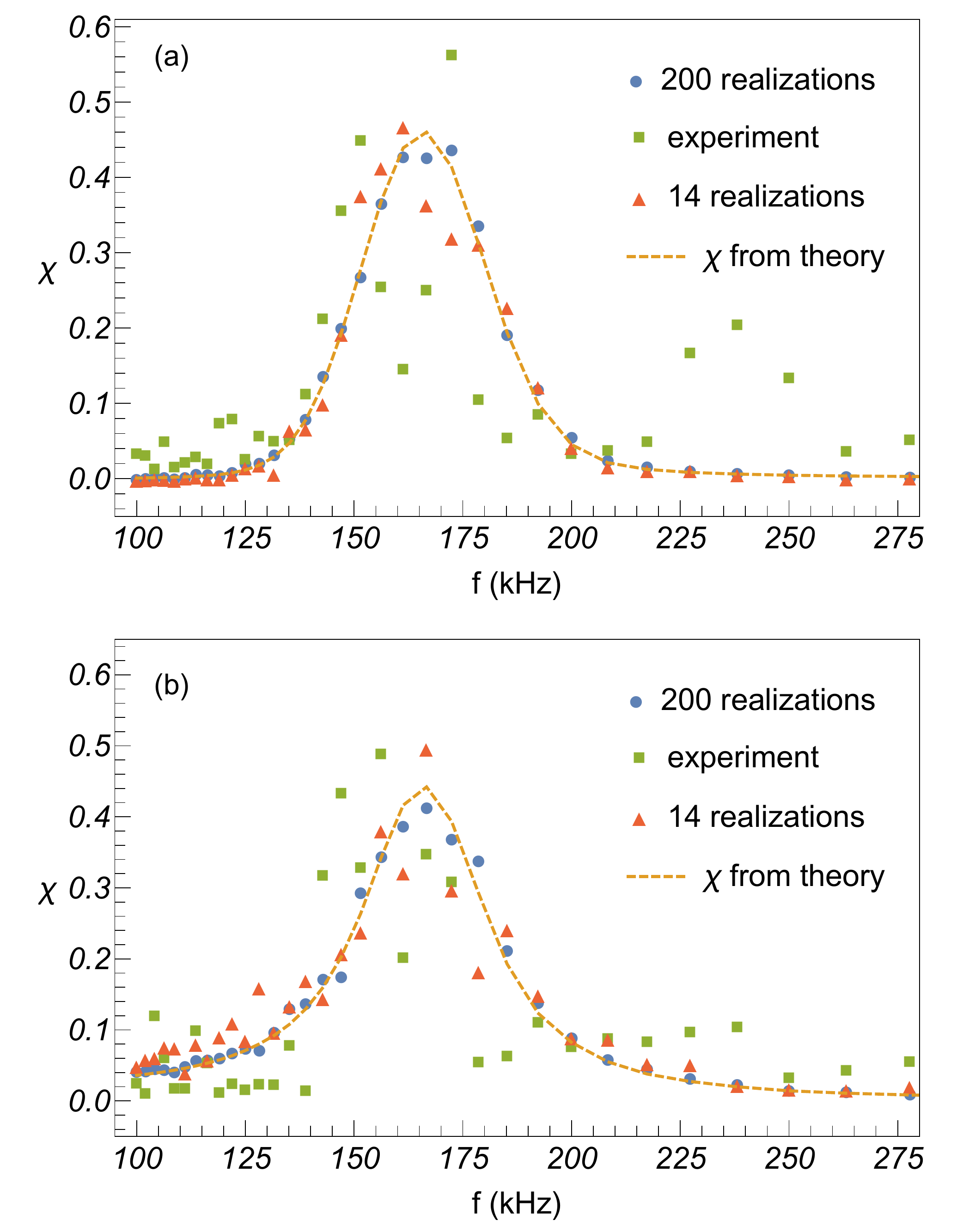}
\caption{
Second-order correlation function $\chi^{(2)}_N$ of a Gaussian (a) and Lorentzian (b) noise power spectral density. Both spectra have the same peak frequency and spectral width. The dashed yellow lines show the theoretical curves given by evaluating Eq.\,(\ref{eq:overlap_spectrum_filter}), while the green squares are the correlation function obtained by experimental measurements with $Q=14$. Instead, the red triangles and blue circles denote the correlation function from numerical simulations of the system dynamics with $Q=14$ and $Q=200$, respectively.
}
\label{fig:GausLor-chi}
\end{center}
\end{figure}

\begin{figure}[h!]
\centering
\includegraphics[width=0.8\textwidth]{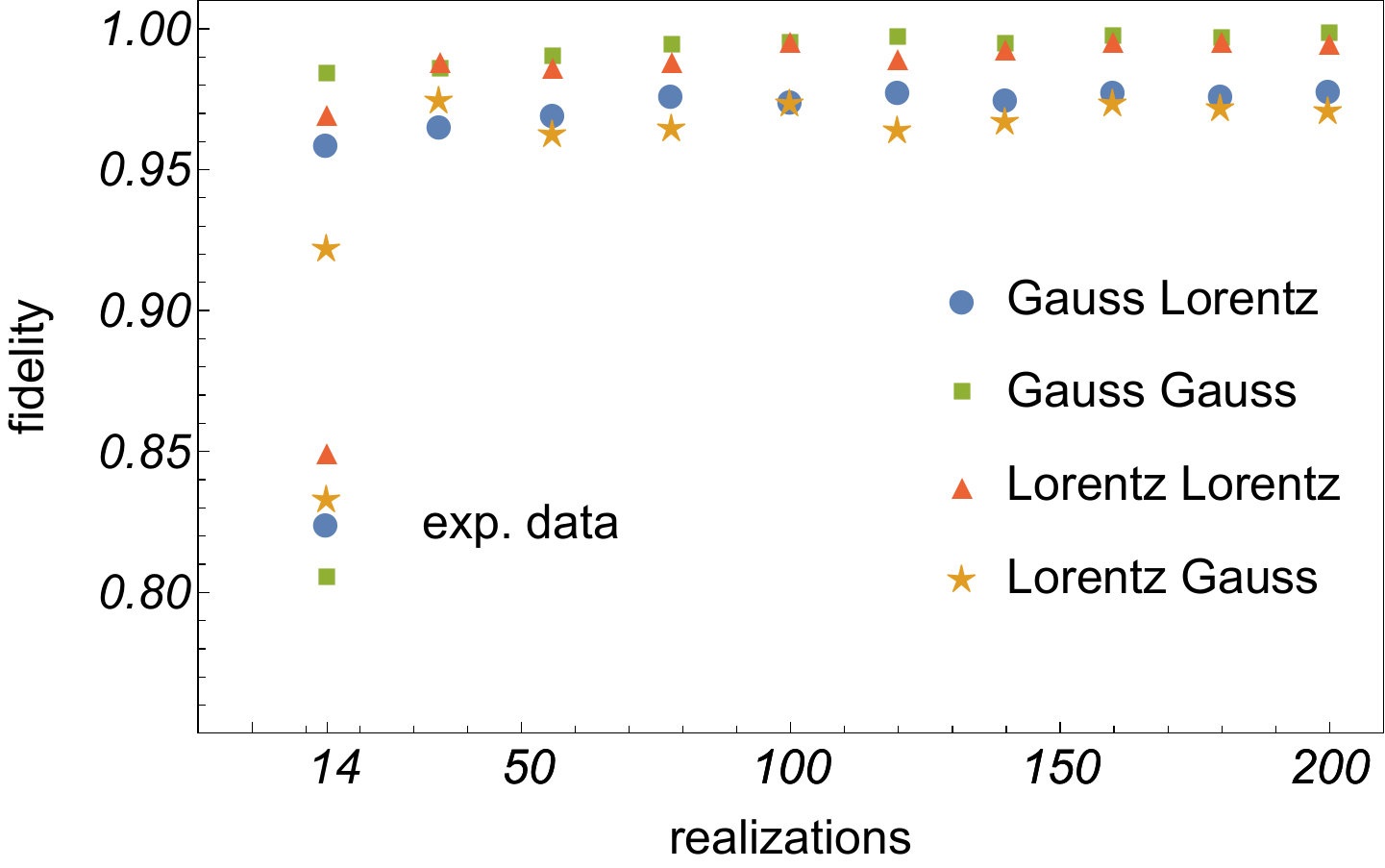}
\caption{
Fidelity $\mathcal{F}_{\chi}$ of the second-order correlation function obtained by comparing the numerical and theoretical data. First, we compare the $\chi^{(2)}_N$ evaluated from the Gaussian power spectral density with the \emph{theoretical} curves given by using Lorentzian (blue circles) and Gaussian (green squares) noise spectra. Then, the $\chi^{(2)}_N$ evaluated from the Lorentzian noise is compared to the theoretical curves from both Lorentzian (red triangles) and Gaussian (yellow stars) noise. Overall, considering the data from the numerical simulations, the fidelity increases with the number of realizations $Q$. Secondly, all plotted data points obtained by comparing $\chi^{(2)}_N$ from the same spectra are much closer to $1$ than the mixed ones. Thus, from the results of the numerical simulations, we can distinguish between Gaussian and Lorentzian spectra. The lower data points on the left corner, referring to $Q=14$, are obtained by directly using the experimental data.
}
\label{fig:GausLor-fidelity}
\end{figure}

\begin{figure}[h!]
\begin{center}
\includegraphics[width=0.76\textwidth]{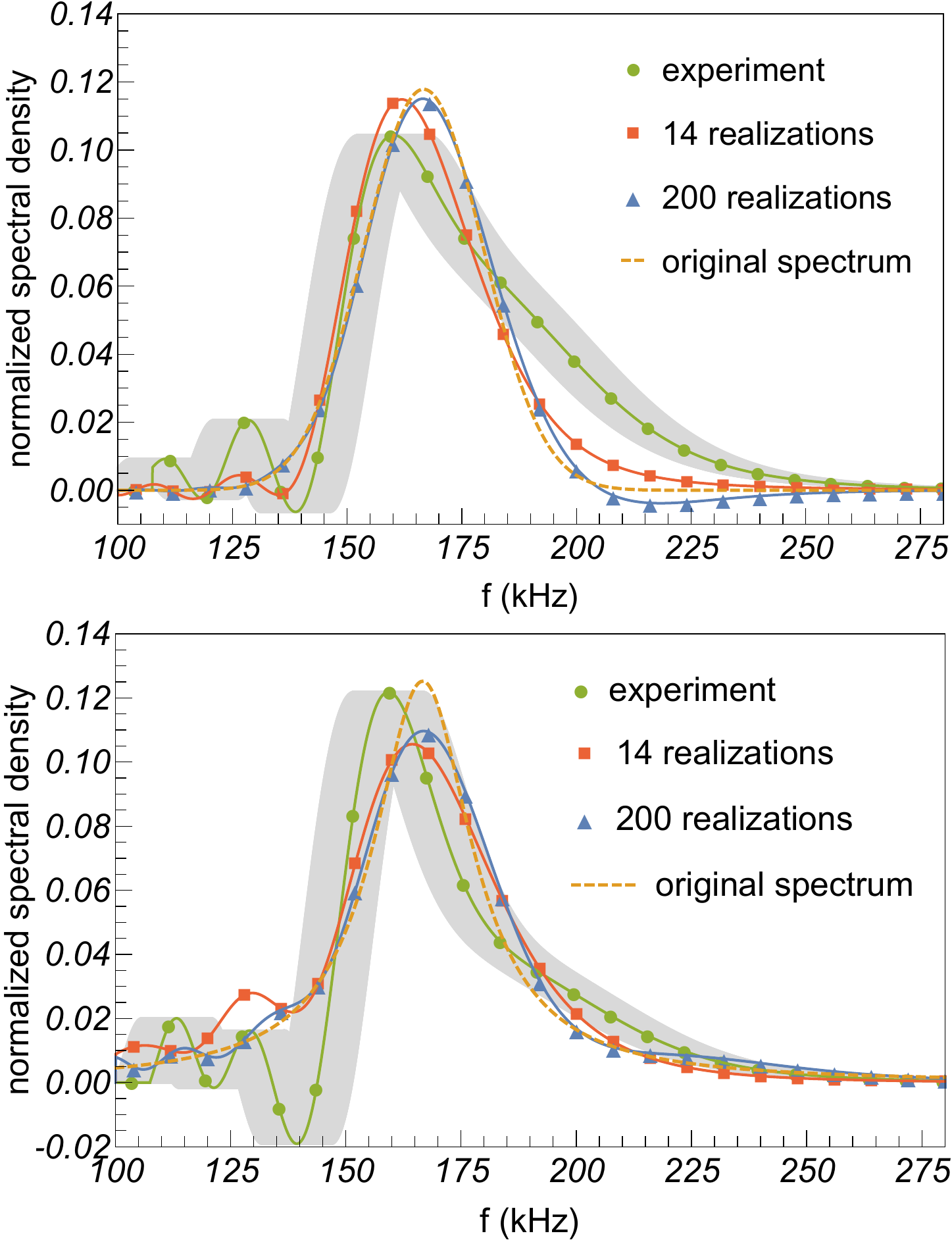}
\caption{
Reconstructed power spectral density for a Gaussian (a) and Lorentzian (b) noise. Both spectra have the same peak frequency and spectral width, and they were reconstructed from the data presented in Fig.\,\ref{fig:GausLor-chi}. The green curves with circle markers show the reconstruction from the experimental data with $Q=14$, while the shaded gray areas denote the frequency uncertainty due to the finite time duration of the laser pulses that realize the projective measurements. Instead, the dashed yellow line is the original power spectral density of the noise (not taking into account the window function stemming from the finite pulse length), while the red line with square markers and the blue one with triangular markers show, respectively, the spectra reconstructed by using the data obtained from numerical simulations of the system dynamics with $Q=14$ and $Q=200$.
}
\label{fig:GausLor-spectrum}
\end{center}
\end{figure}
\begin{figure}[h!]
\centering
\includegraphics[width=0.8\textwidth]{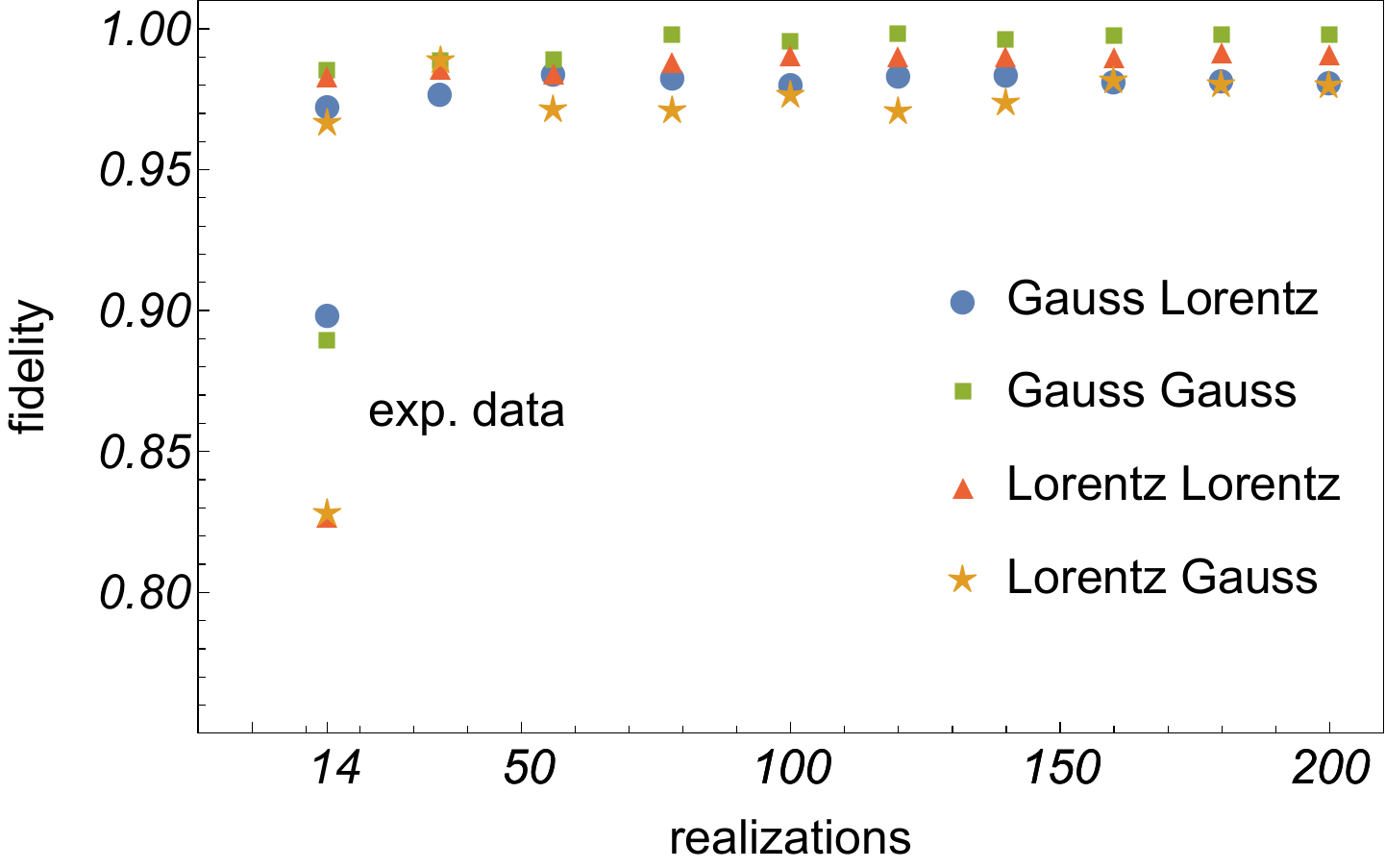}
\caption{
Fidelity $\mathcal{F}_{S}$ obtained by comparing the reconstructed noise power spectral densities with the corresponding \emph{theoretical} expressions. We first compare the reconstructed power spectral density of the Gaussian noise with the theoretical expression of both the original Lorentzian (blue circles) and Gaussian (green squares) spectrum. Then, we relate the reconstructed power spectral density of the Lorentzian noise to both the original Lorentzian (red triangles) and Gaussian (yellow stars) ones. Overall, considering the data from the numerical simulations, the reconstruction fidelity increases with the number of realizations $Q$, and also, in this case, all plotted data points obtained by evaluating the fidelity with the same type of spectra are much closer to $1$ than the fidelities with mixed spectra. Thus, we can distinguish between Gaussian and Lorentzian spectra. The lower data points referring to $Q=14$ compare the reconstructed power spectral densities of noise by directly using the experimental data.
}
\label{fig:GausLor-spectrum-fidelity}
\end{figure}

To prove the protocol in a more realistic scenario, we have repeated the experiment using two \emph{broader} noise power spectral densities, a Gaussian and a Lorentzian one, both centered at $167\rm\,kHz$ with the same width of $50\rm\,kHz$ (FWHM for the Lorentzian, $2\sigma$ for the Gaussian). Both experiments and simulations qualitatively agree with the applied noise spectra, see Figs.\,\ref{fig:GausLor-chi} and \ref{fig:GausLor-fidelity} for the second-order correlation function and Figs.\,\ref{fig:GausLor-spectrum} and \ref{fig:GausLor-spectrum-fidelity} for the power spectral density. In these figures, the experimental data and the theoretical curves are presented together with data obtained from numerical simulations of the quantum probe dynamics. In particular, Fig.\,\ref{fig:GausLor-chi} shows the normalized ($L_2$-norm) second-order correlation functions for both scenarios. Instead, Fig.\,\ref{fig:GausLor-fidelity} shows the overlap (fidelity $\mathcal{F}_{\chi}$ defined below) between the experimental and numerical data points $\chi^{(data)}(\tau_k)$ ($\tau_k$ are the different $K$ values of $\tau $ with $k=1,\dots,K$ for which we repeat the measurements $Q$ times) and the theoretical values $\chi^{(teo)}(\tau_k)$ from Eq.\,(\ref{eq:overlap_spectrum_filter}), i.e.,
\begin{eqnarray}
\mathcal{F}_{\chi}=\sum_{k=1}^K \chi^{(data)}(\tau_k)\chi^{(teo)}(\tau_k)\,.
\end{eqnarray}

Then, we analyze the power spectral densities of the noise fields (see Fig.\,\ref{fig:GausLor-spectrum}). We can reconstruct by means of a filter orthogonalization protocol\,\cite{MuellerSciRep2018} the power spectral density of the noise from the data points $\chi^{(data)}(\tau_k)$, corresponding to the measurement of the overlap between the filter $F_k(\omega)$ and the original spectrum $S^{(orig)}(\omega)$ as given by Eq.\,(\ref{eq:overlap_spectrum_filter}). This technique allows to perform the reconstruction of a given noise spectral density in the $K$-dimensional function space spanned by the filter functions $F_k(\omega)$, by calculating an orthonormal basis $\widehat{F}_k(\omega)$ of this space. In our case this method is even more recommended, since the window-functions of the finite pulse length lead to very broad, comparable to the spectral width of the noise signal, filter functions $F_k(\omega)$ that are non-orthogonal. In a nutshell, we derive the overlap between the filter functions $F_k(\omega)$ in the frequency domain, i.e., we compute the symmetric matrix $A$ whose matrix elements are $A_{kl} \equiv \int_{0}^{\omega_c}F_{k}(\omega)F_{l}(\omega)d\omega$ with $k,l = 1,\ldots,K$ and $\omega_c$ the cut-off frequency. The matrix $A$, being symmetric, can always be diagonalized by means of the transformation $VAV^{\rm T} = \Lambda$, where the matrix $V^{\rm T}$ contains the eigenvectors of $A$ and $\Lambda$ is a diagonal matrix whose elements are the eigenvalues $\lambda_k$, $k=1,\ldots,K$. Thus, we can write down the orthonormal basis functions $\widehat{F}_{k}(\omega) \equiv \frac{1}{\sqrt{\lambda_k}}\sum_{l=1}^{K}V_{kl}F_{l}(\omega)$, $k=1,\ldots,K$  (i.e., $\int_{0}^{\omega_c}\widehat{F}_{k}(\omega)\widehat{F}_{l}(\omega)d\omega = \delta_{kl}$ with $\delta_{kl}$ denoting the Kronecker-delta) and perform the reconstruction of the noise spectral density in this transformed basis. The expansion of the spectral density in the orthonormal basis is
\begin{eqnarray}
S^{(rec)}(\omega) = \sum_{k=1}^K \widehat{c}_k \widehat{F}_k(\omega)\,,\\
\widehat{c}_k(\omega) = \int_0^{\omega_c} S^{(orig)}(\omega)\widehat{F}_k(\omega) = \frac{1}{\sqrt{\lambda_k}} \sum_{l=1}^K V_{kl}\,\chi^{(data)}(\tau_l)\,.
\end{eqnarray}
The main advantage of this reconstruction technique is the robustness against the amount of statistical noise due to the truncation of the orthogonal basis to the dominant eigenvalues, more details can be found in Ref.\,\cite{MuellerSciRep2018}. Notice that the negative values of the reconstructed power spectral densities are an artefact of the reconstruction method within a bigger function space and, thus, they could be set to zero. Here, the spectra, both the original, $S^{(orig)}(\omega)$, and the reconstructed one, $S^{(rec)}(\omega)$, are normalized by the $L_2$-norm, and the fidelity is given by the continuous overlap
\begin{eqnarray}
\mathcal{F}_{S}=\int_{\omega_{min}}^{\omega_{max}} S^{(rec)}(\omega)S^{(orig)}(\omega) d\omega\,,
\end{eqnarray}
where $\omega_{\rm min} = 2 \pi \times 100\rm\,kHz$ and  $\omega_{\rm max} = 2 \pi \times 300\rm\,kHz$ define the frequency range within which we want to reconstruct the power spectral density of the noise.

We apply this estimation technique on both the experimental and the numerical data obtained via simulations with $14$ and $200$ noise realizations. The results are plotted in Fig.\ref{fig:GausLor-spectrum}. We can see that from the experimental data the reconstructed spectrum resembles the original one, although the peak is shifted to lower frequencies and some residual oscillations are present. We believe that this is an effect of the finite time duration of $\tau_{\rm m}$ of the light pulse used to implement the Zeno projective measurements.

We can take this effect into account by considering uncertainty of $\pm \tau_{\rm m}/2$ around each $\tau$, represented by a shaded area in Fig.\,\ref{fig:GausLor-spectrum}. If we take the center of this area, shown as a continuous green line in Fig.\,\ref{fig:GausLor-spectrum}, the reconstruction fidelity increases to $\sim 93\,\%$ for the Gaussian spectrum and to  $\sim 95\,\%$ for the Lorentzian one. In a future experiment, this source of error could be reduced by a combination of stronger lasers (thus, faster projective measurements) and a correction factor in the theoretical treatment of the protocol.

The numerical results, instead, allow for a high-fidelity reconstruction of the noise power spectral densities already for $Q=14$. Moreover, the two spectra can also be distinguished from both the second-order correlation function and the reconstructed spectrum since the fidelities evaluated with the same original and reconstructed spectrum are higher than those evaluated with crossed spectra (see Figs.\,\ref{fig:GausLor-fidelity} and \ref{fig:GausLor-spectrum-fidelity}). In this regard, it is worth noting that the overlap of the original Lorentzian and Gaussian power spectral densities is very high ($97.7\,\%$) and, thus, very precise sensing is required to distinguish the two shapes, which may explain the poor performances obtained with experimental data. Taking around $Q=100$ realizations, we have reconstruction fidelities of well above $99\,\%$.

\section{Conclusion and outlook}
\label{section-conclusions}

In conclusion, we have presented and experimentally demonstrated a new method, based on the stochastic quantum Zeno effect, for estimating the power spectral density of an unknown transverse noise field. This method uses repeated projective measurements together with a control signal on the transverse field to correlate the final survival probability and the noise spectrum via the function $\chi^{(2)}_N$. From the data obtained from a collection of controls we can estimate the noise spectrum via an orthonormalization procedure applied to the filter functions. In the case of a Lorentzian and Gaussian spectrum, we can experimentally reconstruct them with fidelity as large as $80\,\%$ (Lorentzian) and almost $90\,\%$ (Gaussian). This fidelity increases to more than $99\,\%$ in numerical simulations. The latter allows distinguishing the two different spectra.

As an outlook, by enhancing the long-period-stability of the experiment, we aim at increasing the number of repetitions $Q$ while reducing the measurement spread, to improve the quality of the statistics. The numerical analysis shows that already a factor of five would be enough to substantially improve the fidelity of the protocol, and distinguish a Lorentzian power spectral density from a Gaussian one with the same peak position and width. As already mentioned, the finite time duration of the projective measurements is a considerable source of error, which could be reduced in the future by higher laser power and a correction term in the theoretical treatment.

From a purely theoretical point of view, a major step forward could be a better design of the control pulses adopted to reconstruct the noise power spectral densities, which can also be changing in time\,\cite{MuellerSciRep2018}.
Finally, one could increase the reconstruction fidelity by experimentally implementing entangled probes or, as an alternative, by using feedback control and machine-learning enhanced reconstruction methods.

\section*{Acknowledgments}
The author gratefully acknowledges Elisabetta Paladino and Giuseppe Falci for fruitful discussions and comments, and Massimo Inguscio for continuous inspiration and support. H.-V.D.\,and S.G.\,have equally contributed to this work from the experimental and theoretical side, respectively. S.G., N.D.P., and F.C.\,were financially supported from the Fondazione CR Firenze through the project Q-BIOSCAN, PATHOS EU H2020 FET-OPEN grant no.\,828946, and UNIFI grant Q-CODYCES. M.M.\,acknowledges funding from the  EC  H2020  grant 820394 (ASTERIQS).

\section*{Appendix A. Derivation of equation (\ref{chi})}

In this appendix we provide more details on the derivation of Eq.\,(\ref{chi}). By introducing the piecewise constant function $\widetilde\Omega_c(t)$, modeling the sequence of constant control pulses between each couple of projective measurements, the cross-term $P_{\rm cn}$ of Eq.\,(\ref{pcn}) can be written as
\begin{equation*}
P_{\rm cn} = \exp\left[-2\int_{0}^{N\tau}\widetilde\Omega_{c}(t)\Omega_{n}(t)dt\right],
\end{equation*}
with $N$ total number of measurements separated by the time interval $\tau$. In the hypothesis of weak noise, $P_{\rm cn} \approx P/P_{\rm c}$, as explicitly shown in the main text. This means that one has direct access to $P_{\rm cn}$, by measuring the survival probability $P$ at each repetition of the sensing protocol. Then, by averaging the square of the logarithm of $P_{\rm cn}$ over a sufficiently large number of sensing protocol realizations, Eq.\,(\ref{chi}) is obtained. Formally,
\begin{eqnarray*}
    \frac{1}{4}\langle\ln^{2}(P_{\rm cn})\rangle = \left\langle\left[\int_{0}^{N\tau}\widetilde{\Omega}_c(t) \Omega_n(t)dt\right]^2 \right\rangle\nonumber \\
    = \left\langle\int_{0}^{N\tau}\int_{0}^{N\tau} \widetilde{\Omega}_c(t)\widetilde\Omega_c(t')\Omega_n(t)\Omega_n(t')dt dt'\right\rangle\nonumber \\
    = \int_{0}^{N\tau}\int_{0}^{N\tau}\widetilde{\Omega}_c(t)\widetilde{\Omega}_c(t') \left\langle\Omega_n(t)\Omega_n(t')\right\rangle dt dt'\,.
\end{eqnarray*}
We have thus proved that $\langle\ln^{2}(P_{\rm cn})\rangle$ is proportional to the second-order correlation function $\chi^{(2)}_{N}$ as in Eq.\,(\ref{chi}).

\end{document}